\documentclass[]{spie}  

\usepackage{amsmath,amsfonts,amssymb}
\usepackage{graphicx}
\usepackage[colorlinks=true, allcolors=blue]{hyperref}

\title{Characterisation of ALPAO deformable mirrors for the NAOMI VLTI Auxiliary Telescopes Adaptive Optics}

\author[a,c]{Jean-Baptiste Le Bouquin}
\author[a]{Jean-Philippe Berger}
\author[a]{Jean-Luc Beuzit}
\author[a]{Eric Cottalorda}
\author[a]{Alain Delboulbe}
\author[b]{Sebastien E. Egner}
\author[b]{Frederic Yves Joseph Gonte}
\author[a]{Sylvain Guieu}
\author[b]{Pierre Haguenauer}
\author[a]{Laurent Jocou}
\author[a]{Yves Magnard}
\author[a]{Thibaut Moulin}
\author[a]{Sylvain Rochat}
\author[a,b]{Christophe Verinaud}
\author[b]{Julien Woillez}

\affil[a]{Univ. Grenoble Alpes, CNRS, IPAG, 38000 Grenoble, France}
\affil[b]{European Southern Observatory, Garching, Germany}
\affil[c]{Astronomy Department, University of Michigan, Ann Arbor, MI 48109, USA}

\authorinfo{Further author information:\\J.B. Le Bouquin: E-mail: jean-baptiste.lebouquin@univ-grenoble-alpes.fr}

\begin{document} 
\maketitle

\begin{abstract}
The Very Large Telescope Interferometer Auxiliary Telescopes will soon be equipped with an adaptive optics system called NAOMI. The corrective optics deformable mirror is the commercial DM241 from ALPAO. Being part of an interferometer operating from visible to mid-infrared, the DMs of NAOMI face several challenges (high level of reliability, open-loop chopping, piston-free control, WFS/DM pupil rotation, high desired bandwidth and stroke). We here describe our extensive characterization of the DMs through measurements and simulations. We summarize the operational scenario we have defined to handle the specific mirror properties. We conclude that the ALPAO DMs have overall excellent properties that fulfill most of the stringent requirements and that deviations from specifications are easily handled. To our knowledge, NAOMI will be the first astronomical system with a command in true Zernike modes (allowing software rotation), and the first astronomical system in which a chopping is performed with the deformable mirror (5'' sky, at 5~Hz).
\end{abstract}


\section{INTRODUCTION}
\label{sec:intro}

The NAOMI project consists of installing a low order adaptive optics system in each of the four Auxiliary Telescopes (ATs) of the Very Large Telescope Interferometer (VLTI\cite{Haguenauer:2010}) of ESO. The system is composed of (1) a 4$\times$4 Shack-Hartmann visible wavefront-sensor based on a EMCCD ANDOR\footnote{http://www.andor.com} ixon ultra 897, (2) a DM241 Deformable Mirror from ALPAO\footnote{http://www.alpao.com}, and (3) the SPARTA Real time Controller\cite{2012SPIE.8447E..2QS} of ESO running at 500Hz. The NAOMI systems are currently being successfully evaluated in a end-to-end bench mimicking the AT interfaces. The commissioning will take place in November 2018 at the VLTI. More information about the system can be found in the contribution by Frederic Gonte in this proceeding.

The main peculiarity of NAOMI is the choice of a Deformable Mirror (DM) with 241 actuators while the telescope wavefront-sensor has only 4$\times$4 slopes measurements. It arises from the requirement on the pupil size (28mm), and the ambition to perform all fast actuations (turbulence, tip-tilt, 5'' chopping) with a single device. This choice creates a large non-controlled space that should be thoroughly characterise to ensure it meets, and maintains, the requirement of flatness. On the other side, this choice allows to control the DM in true Zernikes, which can be rotated numerically to compensate for the telescope azimuth axis.

This paper presents the characterization of the NAOMI DMs. Section~\ref{sec:dm_calibration} introduces the DM calibration bench and the routine measurements it provides. Section~\ref{sec:modes} details the construction of the modal base of the NAOMI control. Section~\ref{sec:stability} demonstrate that the DMs meet the stability requirements in term of flat, gain and creep. Section~\ref{sec:dynamical} describes the dynamical performance and the multi-stepping scheme used to overcome the DM resonance. Finally, section~\ref{sec:impact} summarizes the performances and the impact of the DM241 on the operation of the NAOMI system.

\section{DM CALIBRATION}
\label{sec:dm_calibration}

A calibration bench is part of the deliverable of the NAOMI project. It will be installed in Paranal observatory in order to re-calibrate the DMs if necessary. The basic concept of the bench is to image the DM surface on a high resolution wavefront sensor. The goal of the bench is to measure the influence function of all actuators at high spatial resolution in order to re-calibrate the mode-to-command matrix, and to re-calibrate the best flat of the membrane.

\subsection{NAOMI calibration bench}
\label{sec:bench}

\begin{figure} [ht]
\begin{center}
\includegraphics[width=0.8\textwidth]{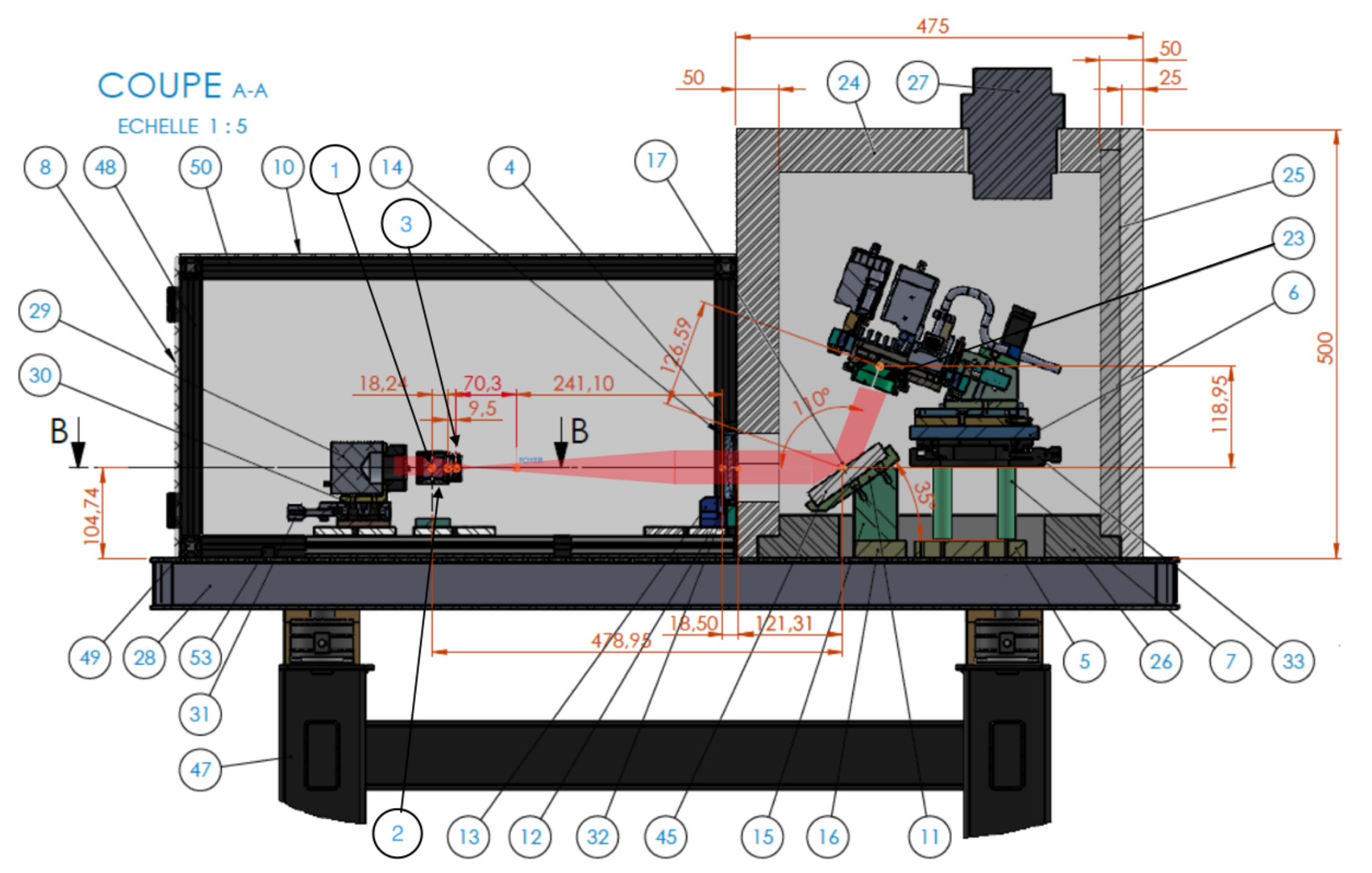}
\end{center}
\caption{\label{fig:bench} Sketch of the NAOMI calibration bench.}
\end{figure}

An overall sketch of the bench can be found in Figure~\ref{fig:bench}. The light source is a multimode fiber plugged to a powerful red visible LED (1). This beam is collimated with the L1 lens. The beam is redirected toward the CO through a 50/50 beam-splitter cube (2). The beam is expanded to 49mm with an afocal system L2/L3 to illuminate properly the DM (3 and 32). The beam is redirected toward the DM inside the CO with a flat folding mirror M1 (17). The beam reflects in the DM at normal incidence (23). The beam is compressed back by the afocal system L2/L3 and is then transmitted though the beam-splitter. The wavefront is measured by a 128$\times$128 Shack Hartman wavefront-sensor from Imagine Optics (29), located in the conjugated plan of the DM. The 14.6mm aperture size of the wavefront-sensor corresponds to 48.2mm in the DM, which is enough to cover the entire 40mm observable pupil. In DM space, the sampling corresponds to 0.38mm/sub-aperture, that is 6.6 sub-apertures/actuator. The 40mm pupil is sampled by 105$\times$105 sub-apertures.

An insulated thermal box (24) with an air air cooling Peltier system allows to perform tests and calibrations between 5\,deg and 2\,deg. The DM mount is thermally insulate with the optical breadboard with the use of low thermal insulator pods (7).

The bench is entirely controlled with a suite of Matlab routines. The software can set the temperature, align the DM, record the Influence Function of each actuator, and close the loop toward the best flat. It then creates automatic reports about the DM health check (dead actuator, flat quality, stroke of modes...) and builds the mode-to-command matrix which can be directly input into the NAOMI Real Time Controller.

\subsection{Piston-included Influence Functions}
\label{sec:ifm}

The Influence Functions are measured by looping on actuators sequentially, and recording two push-pull for each of them. The raw influence function is the difference of the phase screens recorded in push and pull. A typical calibration sequence for the 241 actuators last 30min.

\begin{figure} [ht]
\begin{center}
\includegraphics[width=0.95\textwidth]{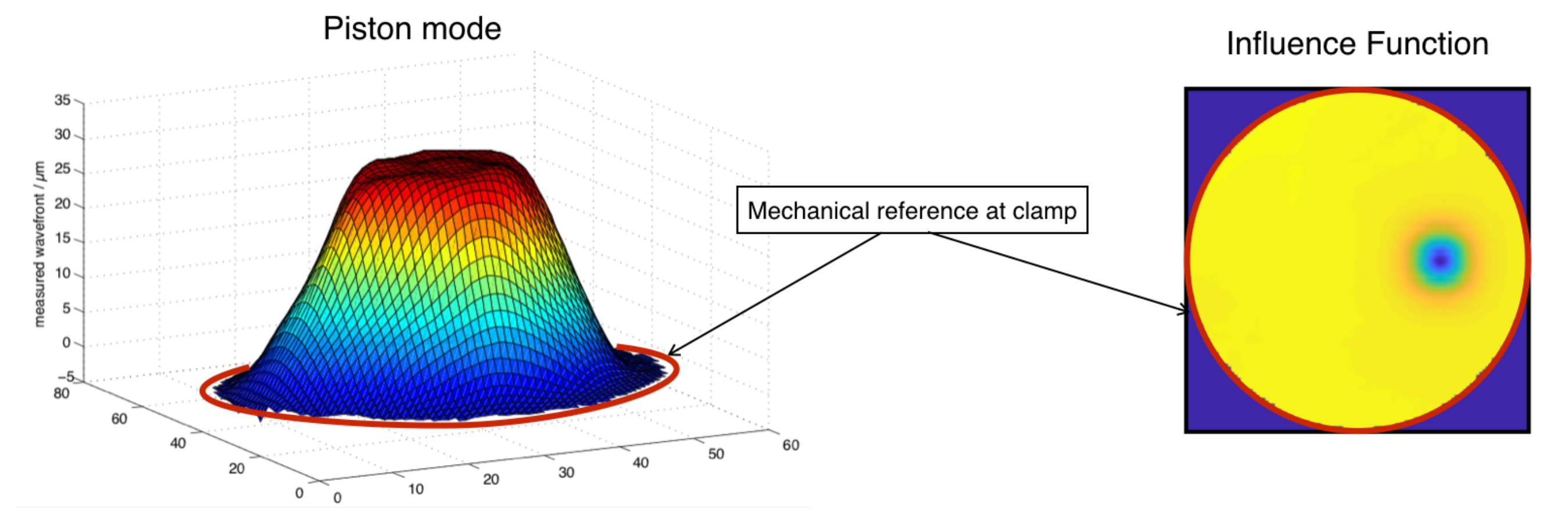}
\end{center}
\caption{\label{fig:IF} 
Measurement of the 'piston' mode (left) and of a single influence function (right) in the calibration bench. The edge of the membrane is a mechanical reference, as demonstrated by the zero derivative in the piston mode.}
\end{figure}

The outer circle of the wavefront ($>50\mu$m) always shows a zero derivative, even when pushing many actuators (see Figure~\ref{fig:IF}). This outer circle defines a mechanical, absolute reference for the wavefront. Consequently, we post-process the raw influence functions to force the outer circle to be at zero-phase. It requires to remove a small global tip and tilt, and to recover the global offset. The global offset obviously corresponds to the fact that the WF is unable to measure the absolute phase. The tip and tilt mostly come from the residual turbulence in the bench, and can be brought to zero if needed by averaging a large number of push-pull.

The post-processed influence functions measured by the calibration bench are therefore absolute phases, meaning that they include the global phase offset, generally omitted in such calibration.

\subsection{DM flat}
\label{sec:flat_calib}

Once all the influence functions have been measured, the software of the bench computes a zonal control matrix. The loop is then closed toward the best flat for about 2min, with a small gain. This allows to flatten the membrane typically at the level of 5nm wavefront-rms. The intrinsic aberration of the bench are mostly of low orders. They are measured by recording a reference phase screen with a flat mirror in the bench, prior to installing the DM in the bench.

\section{THE NAOMI MODES}
\label{sec:modes}

\begin{figure} [ht]
\begin{center}
\includegraphics[width=0.7\textwidth]{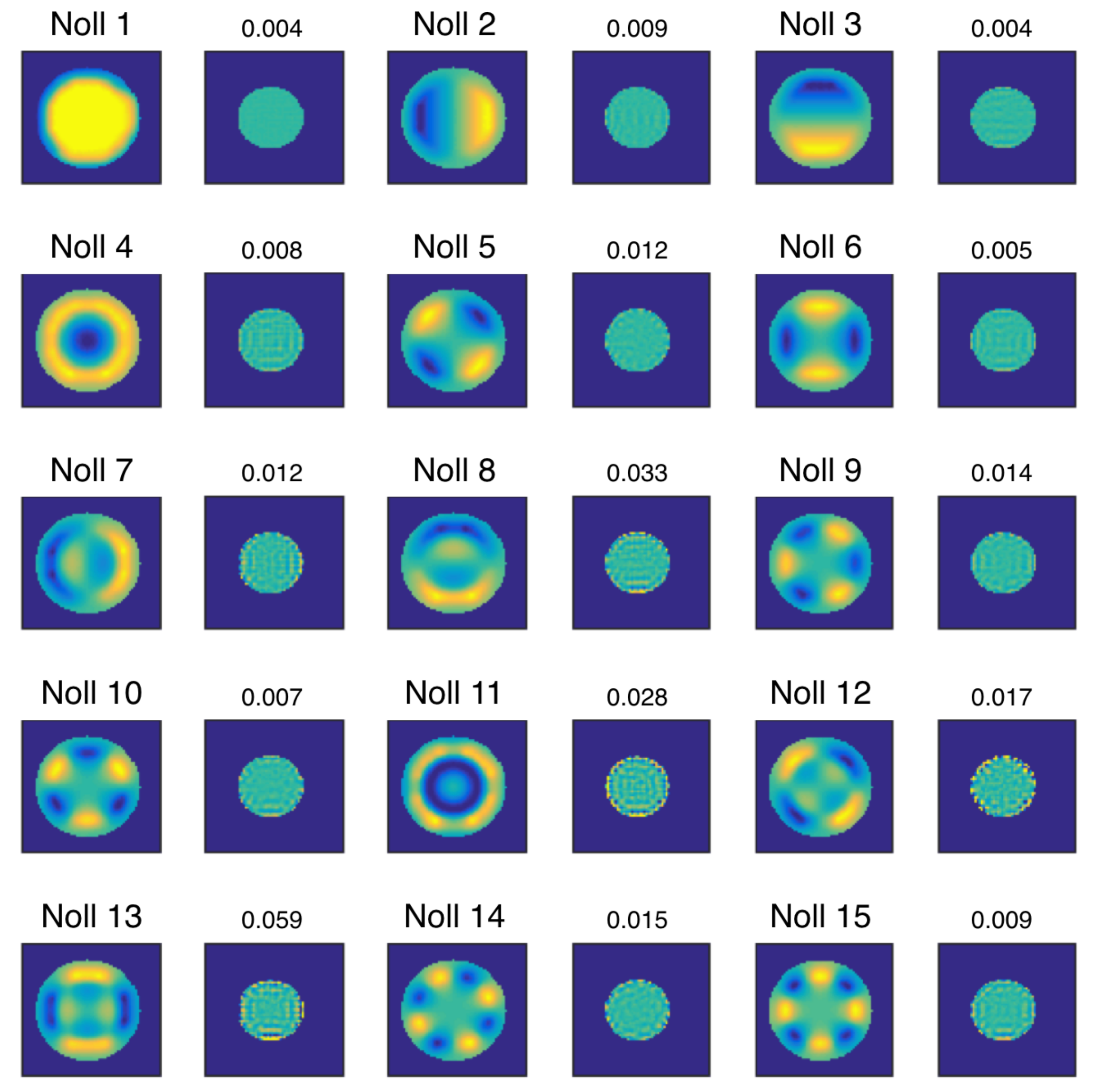}
\end{center}
\caption{\label{fig:NAOMI_mode} 
The 15 NAOMI modes measured across a clear aperture of 48mm, and their residuals to perfect Zernikes across the 28mm NAOMI pupil. The residuals are defined in fraction of the command (e.g $\mu$m  wavefront RMS for a requested amplitude of 1\,$\mu$m wavefront RMS).}
\end{figure}

The sub-aperture of the telescope wavefront-sensor being much larger than the Influence Functions of the DM, it is unrealistic to build a control matrix from a classical zonal interaction matrix. Noise propagation would be very unfavorable. Rather, it has been decided that the Real Time Controller will directly control the DM in Zernikes modes. Numerical experiment demonstrated that the 4$\times$4 wavefront sensor of NAOMI can efficiently control the first 15 Zernikes, that is up to $l=4$ included. The NAOMI modes are shown in figure~\ref{fig:NAOMI_mode}.

\subsection{Building the modes}

In phase space, the NAOMI modes are the first 15 Zernikes modes, defined over the 28mm pupil. They have zero net piston, except mode 1. In the control system, the modes are defined by a mode-to-command matrix $M_m^a$. The basic concept is to build this matrix from accurate and piston-included measurements of the Influence Function of the individual actuators:
\begin{equation}
  \label{eq:modes}
  M_m^a = \mathrm{inv}(\,IF_a^{xy}\,) \;\times\; Z_m^{xy}
\end{equation}
where:\begin{itemize}
\item $xy$ denotes the sub-apertures of the high-resolution 128$\times$128 Shack-Hartman, over the 28mm pupil.
\item $IF_a^{xy}$ is the Influence Function of actuator $a$, including the piston term.
\item $Z_m^{xy}$ is the theoretical Zernike mode $m$.
\item $M_m^a$ is the command vector corresponding to the mode $m$.
\end{itemize}
For the record, the modes are ordered with their Noll number.

\subsection{Filtering unseen actuators}
\label{sec:filtering}

Many actuators $a$ are physically outside the 28mm NAOMI pupil, with some of them having virtually no influence inside the pupil. It was first first envisioned to control these actuators via 'slaving'. However, we found that properly filtering the low-eigenvalues in $\mathrm{inv}(\,IF_a^{xy}\,)$ gives identical results, while being trivial to implement. The choice of the best number of eigenvalues is a compromise between fitting error (difference between desired Zernike and actual shape of DM) and stroke (amplitude of Zernike when first actuator saturates).

We found that the optimal conditioning of $\mathrm{inv}(\,IF_a^{xy}\,)$ is about the same for all Zernike mode $m$. However, it is slightly different from one DM to the next. But this fine-tuning if not necessary to meet the specifications of NAOMI. All-together, we keep 140 eigenvalues out of the 241 Influence Functions. This number matches closely the fraction of actuators inside the 28mm pupil.

\subsection{Zero mean command}
\label{sec:zero_mean_command}

In the course of the dynamical characterization, we discovered that the DMs have a resonance at 500Hz excited by the mean of all commands sent to the actuators (see section~\ref{sec:dynamical}). Consequently, we decided to impose that the sum of all actuators shall be null for all NAOMI modes. In practice, this constrain is implemented as an additional 'phase pixel' in the matrix inversion of equation~\ref{eq:modes}. Let's note $p$ the index of this additional entry in the $xy$ space. The corresponding 'measure' is $IF_a^{p} = 1$, and the corresponding constrains is $Z_m^p = 0$.

\begin{figure} [ht]
\begin{center}
\includegraphics[width=\textwidth]{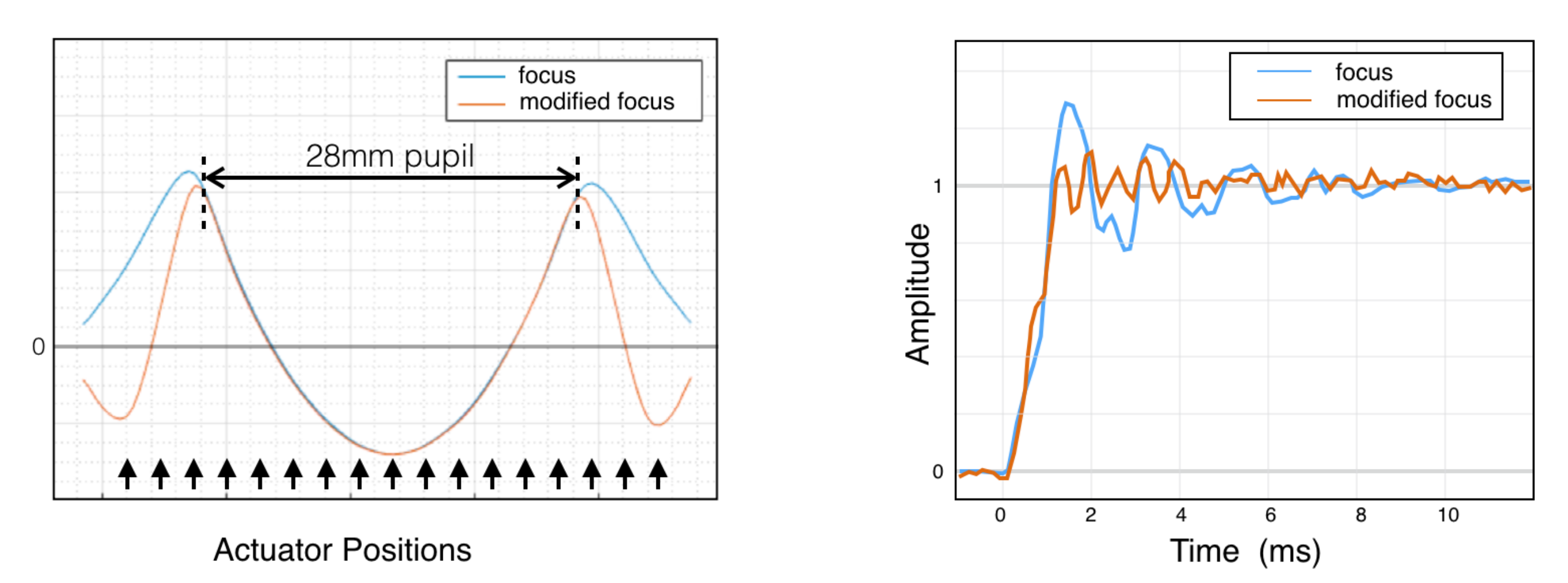}
\end{center}
\caption{\label{fig:focus_mod} 
Left panel: shape of the focus mode without (blue) and with (orange) the additional constrain that the sum of all actuator shall be zero. Right panel: step response of these two modes, as measured by the Keyence position sensor when located in the middle of the pupil.}
\end{figure}

By definition of the Zernikes, all modes except Noll~1 have zero mean phase across the 28mm pupil. But there is a major difference on how this translates into the mean command between radial ($l=0$) and non-radial ($l\neq0$) modes. Thanks to point-symmetry, non-radial modes respect the additional constrain at virtually no cost. But radial modes (piston, focus, spherical) have a significant mean command because the actuators at the edges are all pushing in the same direction. Their stroke is reduced by about 20\% when considering the additional constrain. It remains sufficient to meet the NAOMI requirement. Interestingly, figure~\ref{fig:focus_mod} shows how the matrix inversion naturally uses the unseen actuators to obtain a null sum of all actuators while still maintaining a zero mean piston across the 28mm pupil.

\section{STABILITY OF CALIBRATION}
\label{sec:stability}

The recalibration of a DM requires to dismount it from the telescope. It will be done only few time a year, and ideally even less. The overall stability of the DM calibration in operational environment is therefore of paramount importance.

\subsection{Effect of ambient temperature}
\label{sec:gain}

\begin{figure} [ht]
\begin{center}
\includegraphics[width=0.7\textwidth]{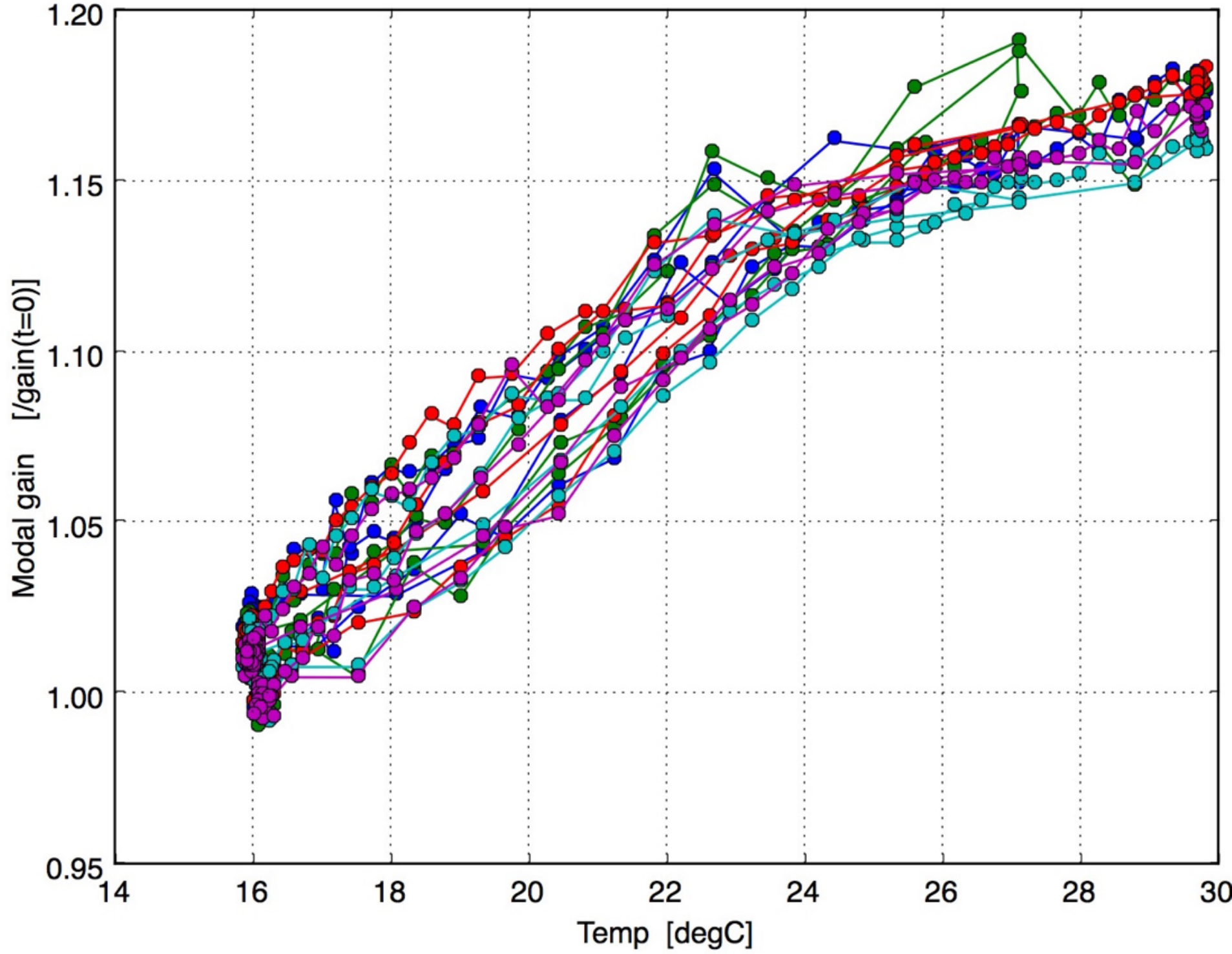}
\end{center}
\caption{\label{fig:Modal_gain} 
Gain for 15 modes versus the ambient temperature, normalized to the gain measured at 16 degC.}
\end{figure}

The standard ALPAO DMs are known to become stiffer at lower temperature. Thanks to the thermal capability of our bench, we can measure the gain of the NAOMI modes at different temperatures. Figure~\ref{fig:Modal_gain} shows the results for temperatures between 16 and 30deg. The same results apply in the operational range of NAOMI (5 to 15deg).

Most interestingly, the evolution of the gain is the same for all measured modes. The trend is +1.5\%/deg, the gain being reduced when considering lower temperature. The NAOMI controller has a look-up table to update the global loop gain depending on the ambient DM temperature. We have not seen any peculiar difficulty due to this modification of the gain.

The reduction of the stroke at low temperature, by the same amount of the gain, was more worrisome. ALPAO can deliver DMs with a 'high stroke' option. Those would have easily matched the specifications, but at the expense of a possibly unacceptable increase of the rise time. This trade-off stroke/speed was the main point of discussion, but the delivered DMs match the demanding specifications of NAOMI (see section~\ref{sec:performances} for an overview).

Finally, the best-flat command is also affected by the temperature. Fortunately, this is a second-order effect because the different actuators have a similar dependency with temperature. We measured that a change of 10deg of ambient temperature increases the wavefront rms from 25nm to 80nm, even after filtering the 15 modes seen by the telescope wavefront-sensor. In order to match the NAOMI specification of 30nm rms, we implemented a look-up table with four different flats for 5, 10 and 15deg.

\subsection{Temperature increase from current in actuators}
\label{sec:temperature_actuator}

The ALPAO DMs are driven by magnetic actuators, which are known to generate more heat than piezo-actuators. We measured an increase of +2deg in the DM housing when using the DM in our worst operational scenario (about 0.5 of the stroke on all actuators), with a typical time constant of 15min to warm-up and cool down. The temperature of the membrane itself is supposed to increase by less than 2deg, since it is physically detached from the coil which generates the head. All together, the impact on NAOMI is considered negligible.

\subsection{Long-term stability of uncontrolled high orders}
\label{sec:flat}

\begin{figure} [ht]
\begin{center}
\includegraphics[width=0.75\textwidth]{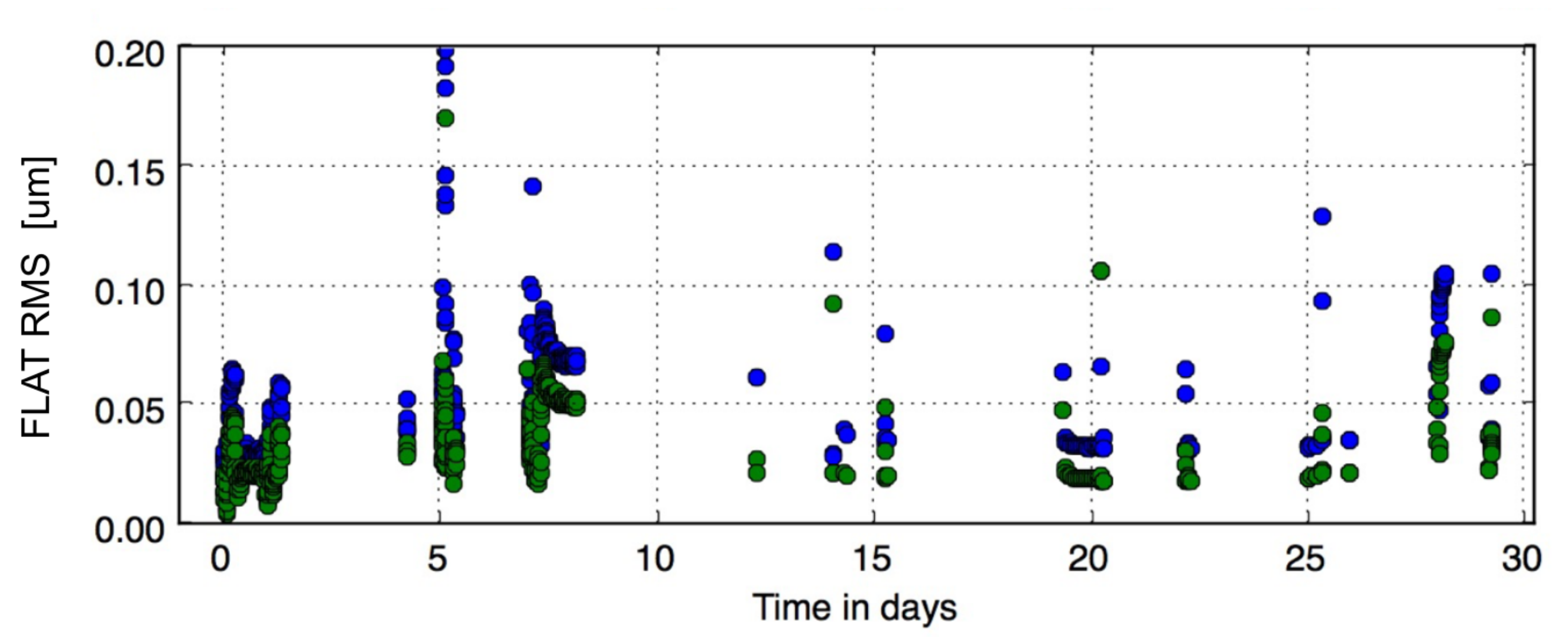}
\end{center}
\caption{\label{fig:Flat_rms} 
Residual wavefront, after correcting 6 modes (blue) and correcting 15 modes (green). The DM flat and mode-to-command has not been re-calibrated during the entire time sequence. The larger residuals measured during day 8 and day 28 correspond to temporary modification of the temperature by up to +10deg.}
\end{figure}

In the course of the development of the bench, we recorded measurements of reflected wavefront with the DM setup in flat command in order to assess the long term stability. Results are shown in Fig.~\ref{fig:Flat_rms}. The low orders (piston, tilts, focus, astigmatism) were not kept under control because of frequent realignments and upgrade of the bench. Assessing the stability of these low order is not critical because they are seen and corrected by the NAOMI wavefront-sensor at the telescope during operation.

We found that the high-order variability is lower than 50nm rms-wavefront over 30 days, after filtering the 15 NAOMI modes. The DM recovers a correct high-order flatness after a temporary warm-up of 10deg over several hours.

\subsection{Creep}
\label{sec:creep}

The technology of the DM241 is subject to creep effect\footnote{ALPAO now offers a new technology not subject to creep, but this option was non-standard at the time of the DM selection and was therefore not considered in the baseline.}. The creep is understood as a slow relaxation of the material used as springs. When compressed/expand, the material slowly modify their internal structure, so that the DM shape slowly evolves under constant command. As far as we could measure, this creep is fully reversal and non-damaging. At first order, the creep is proportional to the applied modification: changing the focus term creates a creep mostly in focus, with less than \%1 of leak in the others Zernikes. However, the amplitude is significant and the time constant is long. The creep deformation  typically amounts for 5\% of the applied command after 1min, 15\% after 10min, 20\% after 100min, and 25\% after several days. Our measurements are quantitatively compatible with the extensive analysis from Ref.~\citenum{Bitenc17}.

The creep effect is completely negligible at the loop frequency (2ms). Regarding chopping, the creep creates an increase of 1\% of tilt during the 1s open-loop SKY position, which is anyway less than the evolution of the uncorrected turbulence.

However, after a long chopping sequence, the DM creeps towards the mean position, that is half way between the STAR and SKY positions. This could be an issue if the DM is supposed to be kept flat in the following minutes. This effect is dramatically reduced by using a symmetric chopping pattern: the STAR position is at -0.5$\times$offset and the SKY position is at +0.5$\times$offset.

\section{DYNAMICAL PERFORMANCES}
\label{sec:dynamical}

The DM241 technology is known to allow a trade-off between stroke and speed. We therefore spend a significant effort into validating the dynamical performances of the DMs, to ensure an optimal configuration. It eventually leads us to implement a multi-stepped command and to modify the shape of the radial modes outside the useful pupil.

\subsection{Multi-stepping}

ALPAO provide a standard measurement of transfer function and step response of actuators. For all DMs, we obtained complementary measurements of step response of modes. The delivered DMs typically have resonance in the range 900-1000Hz for the actuators, and 700-900Hz for the low-order modes. These resonances all have large amplitudes, in the range +15dB to +25dB. Consequently, it was decided to implement a multi-stepping of the command just before sending it to the DM. The basic principle of multi-stepping is to decompose the input 'steps' send by the Real Time Controller (at 500Hz) into several smaller steps (at 4\,kHz) in order to damp the excitation of high frequencies.

\begin{figure} [ht]
\begin{center}
\includegraphics[width=0.9\textwidth]{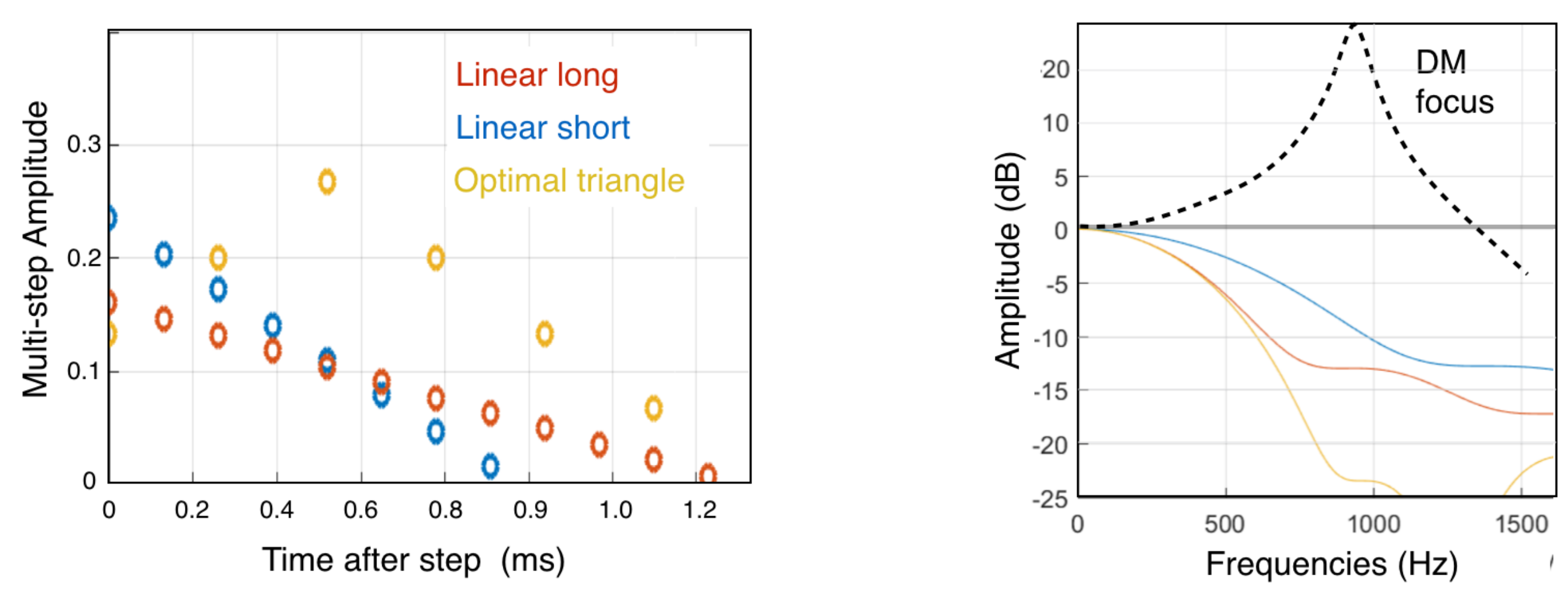}
\end{center}
\caption{\label{fig:multistep} 
Impulse response of three multi-stepping scheme (left), and their corresponding transfer function amplitude (right). The resonance of the DM focus is also represented to show the required damping. NAOMI controller uses the 'optimal triangle'.}
\end{figure}

We numerically tested various scheme. The optimization is a compromise between the attenuation in the resonant frequencies (900Hz) and the phase margin at low frequencies (40Hz). It also corresponds to a compromise between the overshoot and the rise time. Examples are shown in figure~\ref{fig:multistep}. We found that a modified triangular multi-stepping is best suited. This scheme is a variation of the Zero Vibration and Derivative well described in the literature (ZVD, see for instance Ref.~\citenum{William1995}).

In association to our multi-stepping scheme, all ALPAO DMs provide excellent dynamical performances, within the goal requirements of NAOMI, with rise time of 1.1ms or less, and overshoot of 10\% or less, for all modes. Furthermore, current tests show that the initial specifications on resonance amplitude (and overshoot) were conservative. Consequently, we should be able to drive the DM with even faster rise-time.

\subsection{Membrane 'piston' resonance}

We have temporarily installed a Keyence laser position sensor in the calibration bench. According to the documentation, the -3dB bandwidth of the measurement for this used setup is 2.5kHz, therefore we could correctly observe the DM response. Note that the Keyence measure a single point in the membrane surface, and not the entire shape. But it provides an absolute measurement of the displacement.

We verified that the rise time and overshoot measured with the Keyence are in agreement with the ALPAO measurements. This is true for all non-radial modes. However, the piston, the focus and the spherical modes show a new 500Hz oscillation. We conclude that the unexpected 500Hz resonance is related to a global z-motion of the DM surface. This resonance is excited by all radial modes ($l=0$) because the constrain of piston-free over the 28mm pupil makes them non-piston-free over the entire DM surface. The non-radial modes are naturally piston free over all pupil size. This resonance was not detected in the step responses measured by ALPAO because they are based on Shack-Hartman measurements, which filters the global z-motion.

This resonance is impressively suppressed by imposing that the sum of all actuators shall be zero. This is discussed in Section~\ref{sec:zero_mean_command}. The resulting step responses are shown in Figure~\ref{fig:focus_mod}.

\subsection{Power into uncontrolled spatial and temporal frequencies}

The leak of power from the 15 controlled modes toward the higher spatial modes of the DM is a risk for the overall performances. Indeed, every actuator is actually an harmonic oscillator with a resonance frequency near 1000Hz and a very low damping (+15dB). ALPAO made impressive stroboscopic measurements of the dynamical response of the DM. It confirms that the DM shape is not significantly distorted, even dynamically. A tilt mostly excites a tilt resonance of the membrane, a focus mostly excites a focus resonance, and so one.

Even if the excited resonances are out of the control space and thus not critical for the loop stability, they have direct impact on the phase residuals. To quantity this, we estimate the typical size of the step that will be send to the DM. We use a simulation of 4s of turbulences under 1.5" seeing and 10m/s of wind, with wind-shake. We filter this sequence at 100Hz, to account for the limited bandwidth of NAOMI in closed loop (this filtering has negligible impact on results). We consider that the command send to the DM is the running difference of this filtered turbulence over 2ms. Interestingly, we note that all NAOMI modes contribute equally to the typical amplitude of the step command: the turbulence has less amplitude in higher spatial modes but at higher temporal frequencies, thus similar time-derivatives. We then consider the amount of vibration in the measured step responses: typically, 20\% after 2ms, 15\% after 4ms, 10\% after 6ms… We these pessimistic numbers, we found that a mean of 23nm rms-wavefront (median 22nm) of vibration is spread into this uncontrolled space. 

\section{THE DM241 IN NAOMI}
\label{sec:impact}

\subsection{Summary of performances of the delivered DMs}
\label{sec:performances}

Table~\ref{tab:performances} summarizes quantitatively the measured performances of the five NAOMI DM241 delivered by ALPAO, and compare to the specifications.  The modes are the NAOMI modes defined in Section~\ref{sec:modes}. We give the stroke, the linearity and the hysteresis at the lower temperature of the operational range (5deg), which is the most demanding. The stroke at 20deg is about 20\% larger. The dynamical performances make use of the conservative 'optimal triangular' multi-stepping, which contains the overshoot to $<$10\% in all modes and for all command amplitude.

Due to an accident on the DM201, this device was sent back to ALPAO for repair. The final performances are not known at the time of writing, but it seems the repaired version has less stroke than presented here. So far, the tested DMs meet all formal requirements of NAOMI, with the exception of the tip/tilt stroke of the DM199 at 5deg ambient. This DM has nonetheless been accepted because the project had some margin in the stroke and in fitting error which could be relaxed at the time of acceptance thanks to the improved knowledge. Detailed simulation demonstrate that this DM only saturate in the worst operational conditions (5deg, wind shake of telescope structure, chopping on, bad seeing).

   \begin{table}[ht]
\caption{The stroke is defined is reflected wavefront-ptv. The stroke, linearity and the hysteresis are measured at 5deg ambient (worst case). The rise time is measured at 90\%. The rise time and overshoot are similar for the 15 modes, and considers a conservative 'optimal triangular' multi-stepping.}
\label{tab:performances}
\begin{center}       
\begin{tabular}{|l|l|l|l|l|l|l|l|l|l|l|l|}
\hline
DM & Mode 2-3 & Mode 4-6 & Mode 7-15 & Mode & Mode & Actuator & Actuator\\
 \#     & stroke & stroke & stroke & rise-time & overshoot & linearity & hysteresis\\
\hline
 201    & 43\,$\mu$m & 33\,$\mu$m & 15\,$\mu$m & 1\,ms & 5\% & 1.4\% & 2.1\%\\
 200    & 47\,$\mu$m & 32\,$\mu$m & 15\,$\mu$m & 1\,ms & 4\% & 0.5\% & 0.8\%\\
 199    & 33\,$\mu$m & 26\,$\mu$m & 12\,$\mu$m & 1\,ms & 3\% & 0.8\% & 1.9\%\\
 159    & 42\,$\mu$m & 30\,$\mu$m & 14\,$\mu$m & 1\,ms & 4\% & 0.4\% & 1.6\%\\
 153    & 44\,$\mu$m & 31\,$\mu$m & 14\,$\mu$m & 1\,ms & 4\% & 0.3\% & 1.4\%\\
\hline 
Requirement & 40\,$\mu$m & 10\,$\mu$m & 10\,$\mu$m & 1.6\,ms & 15\% & 3\% & 3\%\\
\hline
\end{tabular}
\end{center}
\end{table}

\subsection{Impact on design, operations and performances}
\label{sec:operation}

We here summarize the impact of the choice of the DM241 in the NAOMI design, operation and performances. This is our high-level summary of the interests and risks of using these DMs in an astronomical adaptive optics system.

\begin{itemize}
\item No fast tip-tilt mount, no chopping mount.
\item Control in Zernike modes, allowing software rotation to compensate the telescope azimuth axis.
\item Very fast chopping.
\item Pur-delay significantly smaller than the wavefront-sensor integration time, thus high loop gain.
\item Require a lookup table for modal gain, versus temperature.
\item Require a lookup table for flat of high spatial frequencies, versus temperature.
\item Require a bi-yearly recalibration of high spatial frequencies.
\item Require to chop around mean position.
\item Require to keep the loop closed during several minutes if open-loop high-quality flat is required.
\end{itemize}

\section{CONCLUSIONS}

After careful analysis, and few reasonable modifications of the operational scheme, the ALPAO DM241 meet or exceed all requirements for the NAOMI Adaptive Optics system. To our knowledge, NAOMI will be the first astronomical system with a command in true Zernike modes (allowing software rotation), and the first astronomical system in which a chopping is performed with the deformable mirror (5'' sky, at 5~Hz). Finally, the use of COTS products proved to be a safe route to overcome an accident in the course of the project.

\acknowledgments      
NAOMI is funded by ESO, INSU/CNRS and Universit\'e de Grenoble Alpes. 

\bibliography{report} 
\bibliographystyle{spiebib} 

\end{document}